\documentclass[fleqn,usenatbib,useAMS]{mnras}

\usepackage{graphicx}	
\usepackage{amsmath}	
\usepackage{amssymb}	
\usepackage{multicol}        
\usepackage{bm}		
\usepackage{pdflscape}	
\usepackage[T1]{fontenc}

\usepackage{natbib}
\usepackage{hyperref}
\usepackage[normalem]{ulem}
\usepackage{lscape}
\usepackage{mathtools}
\usepackage{color}
\usepackage{soul}
\usepackage{longtable}
\usepackage{ulem}
\usepackage{makecell}
\usepackage{multirow} 

\usepackage{graphicx}
\usepackage{txfonts}

\title[Energy levels and transitions for Th\,{\sc iii}]{Theoretical study of Th\,{\sc iii} energy levels and transitions for applications to kilonova spectra}

   \author[L. Kitovien\.{e} et al.]
	 {L. Kitovien\.{e},$^{1}$ \thanks{E-mail: laima.radziute@tfai.vu.lt}
	  G. Gaigalas,$^{1}$
		P. Rynkun,$^{1}$
    N. Domoto,$^{2}$
    M. Tanaka,$^{2,3}$
    and D. Kato$^{4,5}$
		\\
	$^1$Institute of Theoretical Physics and Astronomy, Vilnius University, Saul\.{e}tekio Ave. 3, LT-10257 Vilnius, Lithuania\\
	$^2$Astronomical Institute, Tohoku University, Sendai 980-8578, Japan
\\	
	$^3$Division for the Establishment of Frontier Sciences, Organization for Advanced Studies, Tohoku University, Sendai 980-8577, Japan
\\	
	$^4$National Institute for Fusion Science, 322-6 Oroshi-cho, Toki 509-5292, Japan\\
	$^5$Interdisciplinary Graduate School of Engineering Sciences, Kyushu University, Kasuga, Fukuoka 816-8580, Japan}

\date{Accepted XXX. Received YYY; in original form ZZZ}

\pubyear{2025}

\begin{document}
\label{firstpage}
\pagerange{\pageref{firstpage}--\pageref{lastpage}}
\maketitle
 
\begin{abstract}
The neutron star merger is a promising site of heavy element production. By producing heavy elements, the neutron star merger gives rise to a thermal transient called a kilonova. Studying kilonova spectra enables us to quantify the heavy element production.
Among the heaviest elements, doubly ionized Thorium (Th, $Z$=90) is one of the important candidates for producing detectable absorption features in kilonova spectra. This paper investigates the atomic properties of Th\,{\sc iii} to provide energy level and transition data. 
 The multiconfiguration Dirac-Hartree-Fock and
 relativistic configuration interaction methods, 
which are implemented in the general-purpose relativistic atomic structure package {\sc Grasp}2018, 
are used to compute energy levels of the 
$\mathrm{5f6d}$, $\mathrm{6d^2}$, $\mathrm{7s^2}$, $5\mathrm{f^2}$, $\mathrm{6d7s}$, $\mathrm{5f7p}$ and $\mathrm{5f7s}$
 configurations and 
electric dipole transitions between 
states of these configurations.
The accuracy of energy levels is evaluated by comparing it with experimental data 
and with various theoretical methods.
Our calculated energy levels are consistent with the experimental
results with a root mean square (RMS) deviation of 436 cm$^{-1}$. 
The accuracy of transition data is investigated using 
 the quantitative and qualitative evaluation method.
 By performing radiative transfer simulations for kilonova spectra with our transition data, we show that kilonova including Th with a mass fraction of $(3-10) \times 10^{-5}$ can produce Th\,{\sc iii} absorption features around 18,000 \AA.
\end{abstract}


\section{Introduction}

Atomic data of heavy elements are crucially important to understanding spectroscopic features of astronomical objects. 
In particular, recent observations of a neutron star merger and subsequent electromagnetic signal "kilonova" (GW170817/AT2017gfo, \citealt{abbott17}) enabled us to study the heavy element production in the Universe directly.
For AT2017gfo, extensive time series of optical and near-infrared spectra have been obtained \citep[e.g.,][]{chornock17,pian17,smartt17}.
In these spectra, several heavy elements have been identified/proposed, including Sr \citep{watson19,domoto21,gillanders22}, Y \citep{sneppen23}, La \citep{domoto22} and Ce \citep{domoto22,domoto23,tanaka23,gillanders24,gaigalas24} as absorption features and Te \citep{hotokezaka23} as an emission feature.

As a site of heavy element production, 
one of the most critical questions is how heavy elements are synthesized in neutron star mergers.
In particular, the production of actinides ($Z=89-103$) has been of great interest for the nucleosynthesis in neutron star mergers \citep[e.g.,][]{holmbeck19,wanajo22} as well as its connection to the abundances found in metal-poor stars \citep[see, e.g.,][for a review]{cowan21}.
To understand the impact of actinides on kilonova, several works have been performed \citep{fontes20,fontes23,pognan24}.
However, due to a lack of accuracy, these works only focus on the statistical properties of actinides.

Recently, \citet{domoto24} compiled spectroscopically accurate atomic data for Th\,{\sc iii} and suggested that Th\,{\sc iii} lines can show absorption features in kilonova spectra at near-infrared wavelengths.
However, in their work, transition probabilities (and oscillator strengths) of Th\,{\sc iii} lines have been estimated by using the emission line ratios measured in 
the thorium-argon hollow cathode lamp spectrum \citet{engleman03} together with calculations by \citet[see below]{Biemont_2002_MNRAS}.
Thus, the accuracy of the transition probabilities is hard to estimate.
To assess the detectability of Th\,{\sc iii} lines in the kilonova spectra, 
it is essential to provide accurate and systematic transition data, particularly in the infrared wavelength ranges.

Several experimental and theoretical works on Th\,{\sc iii}, are described below.
In the first papers, the authors used the 
Zeeman patterns to analyze the Th\,{\sc iii} spectrum.
\cite{Klinkerberg_1} investigated the patterns of the lines in the wavelengths range 2099.42 - 4555.64 \AA~ 
and reported 50 identified energy levels. 
\cite{Klinkerberg_2} have observed about 200 lines in the wavelength range 2000 - 9000 \AA~ 
and identified 77 energy levels of the $\mathrm{5f,\{7s,7p,6d\}}$, $\mathrm{5f^2}$,  $\mathrm{6d\{7s,7p\}}$, $\mathrm{7p7s}$, $\mathrm{6d^2}$ configurations.

The number of identified lines was increased significantly using sliding spark spectrum by \cite{Wyart}.
 The authors measured the sliding spark spectrum in the 500 - 1500~\AA~range. In this study, 488 lines were classified as
 transitions between 98 new levels and 77 previously known levels.
The interpretation of the $\mathrm{5f^2}$ and $\mathrm{5f6d}$ configurations has been completed and
the main properties of $\mathrm{5f\{6f,8p\}}$, and $\mathrm{6d\{7d,6f\}}$ were described. All of these
configurations have been interpreted using the Slater-Condon theory. 

\cite{Biemont} have measured lifetimes of six levels, 
with the time-resolved laser-induced fluorescence (LIF)
technique.  
The properties of the transitions depopulating 
these levels were computed using
pseudo-relativistic Hartree-Fock (HFR) method \citep{Cowan}
and normalized with the experimental lifetimes. The first set of transition
probabilities have been deduced for Th\,{\sc iii} by these authors. 

Several theory-based studies of this ion 
have been carried out in recent decades. 
\cite{Dzuba} have computed eleven energy levels
 using a combination of second-order many-body
perturbation theory and configuration interaction, which considers
dominating relativistic and correlation effects (CI$+$MBPT). 
Later, \cite{Safronova} have determined energy levels, 
transition properties, and lifetimes with 
configuration-interaction plus all-order (CI$+$all) method.  
This method combines CI and coupled-cluster approaches. 
\cite{ROY201225} have conducted a series of relativistic
spin-free calculations at state-average complete active space self-consistent field (SA-CASSSCF) for energy levels and transition properties. 
The dynamic electron correlation effects have been studied through second-order 
perturbation theory that has been performed at multi-state second-order
multiconfigurational perturbation theory (MS-CASPT2) 
level. In both cases, the spin-orbit treatments at SA-CASSCF
and MS-CASPT2 levels have been performed
using an effective Fock-type one-electron spin-orbit operator. 

The most recent experiment was done by \cite{Redman_2014}. 
They have made precise and extensive observations of a 
thorium-argon hollow cathode lamp emission spectrum in 3500 - 11750~\AA~region, 
using a high-resolution Fourier transform spectrometer. This investigation 
led to new energy levels of doubly ionized thorium, and these levels are referred to in the NIST \citep{NIST_ASD}. 

As described above, despite its importance in kilonova spectra, the number of papers addressing the properties of Th\,{\sc iii} is still limited. 
Also, as seen from the energy level analysis below, the energy information for the configurations chosen by the authors is incomplete, as is the transition information between these levels. 
It is also worth noting that the discrepancy between the calculated energy level and the experimental values is huge in some authors' theoretical works.
Therefore, we have decided to research using the {\sc Grasp}2018 program package.

We have computed 63 energy levels of the 
$\mathrm{5f\{7s,6d,7p\}}$, 
$\mathrm{6d^2}$,
$\mathrm{5f^2}$, $\mathrm{7s^2}$, and $\mathrm{6d7s}$ configurations, 
using multiconfiguration Dirac-Hartree-Fock (MCDHF) and
 relativistic configuration interaction (RCI) methods.  
Transitions of E1 were computed between these levels. 
The {\sc Grasp}2018 package was used for the investigation \citep{grasp2023}.

\begin{table*} 
\setlength{\tabcolsep}{5pt}
\caption{Comparison of the lowest energy levels (in cm$^{-1}$) of ours (RCI),
 \citet{Safronova} (CI$+$all), 
\citet{Dzuba} (CI$+$MBPT), 
\citet{ROY201225} (SA-CASSCF and MS-CASPT2) 
with the experimental \citet{Redman_2014} (Ex.) values for odd configurations.}
\label{Energies_o}
\begin{tabular}{rrrrrrrrrrrrrrrrrrr}
\hline\hline      
\multicolumn{1}{l}{\multirow{2}{0.10cm}{No.~~}}
& \multicolumn{1}{l}{\multirow{2}{0.10cm}{Label}} 
& \multicolumn{1}{c}{\multirow{2}{0.10cm}{$J$}}
& \multicolumn{1}{c}{\multirow{2}{0.10cm}{$P$}} 
& \multicolumn{1}{c}{Ex.} 
& \multicolumn{2}{c}{CI$+$all} 
&& \multicolumn{2}{c}{CI$+$MBPT} 
&& \multicolumn{2}{c}{SA-CASSCF} 
&& \multicolumn{2}{c}{MS-CASPT2}
&& \multicolumn{2}{c}{RCI} \\ 
\cline{6-7}\cline{9-10}\cline{12-13}\cline{15-16}\cline{18-19}
& & & 
&\multicolumn{1}{c}{$E$}
& $E$ & $\Delta E$  
&&$E$ & $\Delta E$ 
&&$E$ & $\Delta E$ 
&&$E$ & $\Delta E$ 
&&$E$ & $\Delta E$ \\
\hline
\noalign{\smallskip}        		
31 & $\mathrm{5f6d~^3P}$ & 0 & - &  11233 &  11766 &   -533 &&      &      &&       &       &&       &         && 11567 & -334  \\
21 & $\mathrm{5f6d~^3D}$ & 1 & - &  7921  &  8260  &   -339 &&      &      &&       &       &&       &         && 8155  & -234  \\
30 & $\mathrm{5f6d~^3P}$ & 1 & - &  11123 &  11564 &   -441 &&      &      &&       &       &&       &         && 11466 & -343  \\
43 & $\mathrm{5f6d~^1P}$ & 1 & - &  20711 &  22733 &  -2022 &&      &      &&       &       &&       &         && 21789 & -1078 \\
3  & $\mathrm{5f6d~^3F}$ & 2 & - &  511   &  189   &    322 &&      &      && 6933  & -6422 &&  1074 & -563    && 469   & 42    \\
6  & $\mathrm{5f7s~^3F}$ & 2 & - &  3182  &  2958  &    224 &&      &      && 10671 & -7489 &&  5048 & -1866   && 3256  & -74   \\
15 & $\mathrm{5f6d~^1D}$ & 2 & - &  6288  &  5797  &    491 &&      &      && 16333 & -10045&&  8162 & -1874   && 6243  & 45    \\
24 & $\mathrm{5f6d~^3D}$ & 2 & - &  10181 &  10458 &   -277 &&      &      &&       &       &&       &         && 10250 & -69   \\
33 & $\mathrm{5f6d~^3P}$ & 2 & - &  13208 &  13513 &   -305 &&      &      &&       &       &&       &         && 13431 & -223  \\
4  & $\mathrm{5f7s~^3F}$ & 3 & - &  2527  &  2436  &     91 &&      &      && 12099 & -9572 &&  6879 & -4352   && 2548  & -21   \\
9  & $\mathrm{5f6d~^3F}$ & 3 & - &  4827  &  4853  &    -26 &&      &      && 9316  & -4489 &&  4044 & 783     && 4810  & 17    \\
11 & $\mathrm{5f6d~^3G}$ & 3 & - &  5061  &  5085  &    -24 &&      &      && 7629  & -2568 &&  5815 & -724    && 5002  & 59    \\
18 & $\mathrm{5f7s~^1F}$ & 3 & - &  7501  &  7609  &   -108 &&      &      && 14709 & -7208 &&  9515 & -2014   && 7630  & -129  \\
28 & $\mathrm{5f6d~^3D}$ & 3 & - &  10741 &  11236 &   -495 &&      &      &&       &       &&       &         && 10858 & -117  \\
35 & $\mathrm{5f6d~^1F}$ & 3 & - &  15453 &  16506 &  -1053 &&      &      &&       &       &&       &         && 15674 & -221  \\
1  & $\mathrm{5f6d~^3H}$ & 4 & - &  0     &  0     &      0 &&    0 &     0&& 5196  & -5196 &&  1713 & -1713   &&    0  & 0     \\
5  & $\mathrm{5f6d~^1G}$ & 4 & - &  3188  &  3207  &    -19 &&      &      && 8975  & -5787 &&  4938 & -1750   && 3040  & 148   \\
14 & $\mathrm{5f7s~^3F}$ & 4 & - &  6311  &  6237  &     74 &&      &      && 16012 & -9701 &&  10806& -4495   && 6199  & 112   \\
19 & $\mathrm{5f6d~^3G}$ & 4 & - &  8142  &  8197  &    -55 &&      &      &&       &       &&       &         && 8024  & 118   \\
23 & $\mathrm{5f6d~^3F}$ & 4 & - &  8981  &  9063  &    -82 &&      &      && 13033 & -4063 &&  7942 & 1039    && 8867  & 114   \\
8  & $\mathrm{5f6d~^3H}$ & 5 & - &  4490  &  4802  &   -312 &&      &      && 9945  & -5455 &&  7176 & -2686   && 4460  & 30    \\
29 & $\mathrm{5f6d~^3G}$ & 5 & - &  11277 &  11456 &   -179 &&      &      &&       &       &&       &         && 11189 & 88    \\
39 & $\mathrm{5f6d~^1H}$ & 5 & - &  19010 &  20144 &  -1135 &&      &      &&       &       &&       &         && 19719 & -709  \\
22 & $\mathrm{5f6d~^3H}$ & 6 & - &  8437  &  8810  &   -373 &&      &      && 14321 & -5884 &&  11552& -3115   && 8272  & 165   \\
\hline
\end{tabular}
\end{table*}

\section{Computational procedure} 
\label{comp_proc}

\subsection{Methods}
\label{sec:method}
The calculations were performed using the {\sc Grasp}2018 package \citep{grasp2023}, which  
is based on the MCDHF 
and RCI methods.
The first method was used to generate numerical representations of 
atomic state functions (ASFs). 
The angular integrations needed for the construction of the functional energy
are based on second quantization in the coupled tensorial
form \citep{angular1,angular2}. 
The Breit interaction, as well as quantum electrodynamic (QED) corrections,
 are included using the second method.
ASFs are transformed from the
$jj-$ to the $LS-$coupling with a unique representation \citep{unic}. 
 The method described by \citep{GAIGALAS_couling} was used to transform the other couplings. 
The quantitative and qualitative evaluation method (QQE)~\citep{Pavel_Ce,Pavel_Pr,Sb-like,Kitoviene_Ge} 
was used to estimate the uncertainty of the calculated line strengths.
More details about these methods can be found in \cite{topical_rev} and \cite{grant}.

\subsection{Correlation inclusions scheme}
\label{sec:Correlation}

For the calculations, the core was assumed to be the 
[Xe]$\mathrm{4f^{14}5d^{10}6s^26p^6}$ shells. 
The radial wave functions of the core orbitals 
were calculated using the Single Reference configuration method and
 the [Xe]$\mathrm{4f^{14}5d^{10}6s^26p^65f7s}$ configuration was selected for it 
like that was done in the works of \cite{radziute20} and \cite{radziute21}. 
The valence orbitals $\mathrm{\{7s,7p,6d,5f\}}$ were computed in the Multi-reference configuration (MR) approach,  
the $\mathrm{5f\{7s,6d,7p\}}$, $\mathrm{6d^2}$, $\mathrm{5f^2}$, $\mathrm{7s^2}$, and $\mathrm{6d7s}$
configurations were selected for the MR set. 
The sets of virtual orbitals were generated with single and 
double electron excitations from the configurations 
in the MR set. Only valence-valence
 electron correlations were included in this step.  
 Radial functions of virtual 
orbitals were computed using the multiconfiguration 
Dirac-Hartree-Fock method. The calculations were carried 
out step by step 
(or layer by layer, keeping the previous layer frozen).  
The four layers of virtual orbitals (up to $\mathrm{{11s,11p,10d,9f,8g}}$)
 were computed until convergence of the energy levels was achieved.   

For further RCI calculations, the configurations were divided into three groups. 
The first group consists of the $\mathrm{5f\{6d,7s\}}$ configurations, 
the second group consists of $\mathrm{6d^2}$, $\mathrm{7s^2}$, and $\mathrm{6d7s}$,
and the third group includes only the $\mathrm{5f^2}$ and $\mathrm{5f7p}$ configurations. 
Core-valence (CV) electron correlations of $\mathrm{5d6s6p}$ shells were included for all three groups.

The MR set for computation of the energy levels of the  $\mathrm{5f\{6d,7s\}}$ 
configurations (the first group) was extended by $\mathrm{7s7p}$ and $\mathrm{6d7p}$ configurations. 
Core-core (CC) of $\mathrm{6p}$ were included in RCI computations.

The energy levels of the $\mathrm{6d^2}$, $\mathrm{7s^2}$, and $\mathrm{6d7s}$ configurations set (the second group) 
were computed using the MR set: 
$\mathrm{6d^2}$, $\mathrm{7s^2}$, $\mathrm{5f^2}$ $\mathrm{6d7s}$, and $\mathrm{5f7p}$ 
and CC correlations of the shells $\mathrm{6s6p}$ were added.

To compute the energy levels of the $\mathrm{5f^2}$ and $\mathrm{5f7p}$ configurations set  (the third group) of 
 $\mathrm{6d^2}$, $\mathrm{7s^2}$, $\mathrm{5f^2}$ $\mathrm{6d7s}$, and $\mathrm{5f7p}$ configurations were used as the MR and CC correlations of the $\mathrm{6p}$ shells were included. 
 
The final results (RCI) are computed in the basis of 6 227 777 configuration state functions (CSF) for even and 3 426 892 for odd parity.  
The final energy levels are given in Tables \ref{Energies_o} and \ref{Energies_e} for odd and even configurations, respectively.   

\begin{table*} 
\setlength{\tabcolsep}{5pt}
\caption{
Comparison of the lowest energy levels (in cm$^{-1}$) of ours (RCI),
 \citet{Safronova} (CI$+$all), 
\citet{Dzuba} (CI$+$MBPT), 
\citet{ROY201225} (SA-CASSCF and MS-CASPT2) 
with the experimental \citet{Redman_2014} (Ex.) values for even configurations.}
\label{Energies_e}                                                        
\begin{tabular}{rrrrrrrrrrrrrrrrrrr}       
\hline\hline      
\multicolumn{1}{l}{\multirow{2}{0.10cm}{No.~~}}
& \multicolumn{1}{l}{\multirow{2}{0.10cm}{Label}} 
& \multicolumn{1}{c}{\multirow{2}{0.10cm}{$J$}}
& \multicolumn{1}{c}{\multirow{2}{0.10cm}{$P$}}
& \multicolumn{1}{c}{Ex.}  
& \multicolumn{2}{c}{CI$+$all}
&& \multicolumn{2}{c}{CI$+$MBPT} 
&& \multicolumn{2}{c}{SA-CASSCF}
&& \multicolumn{2}{c}{MS-CASPT2}
&& \multicolumn{2}{c}{RCI}\\
\cline{6-7}\cline{9-10}\cline{12-13}\cline{15-16}\cline{18-19}
& & & 
& \multicolumn{1}{c}{$E$} 
& $E$ & $\Delta E$ 
&& $E$ & $\Delta E$ 
&& $E$ & $\Delta E$ 
&& $E$ & $\Delta E$ 
&&$ E$ &  $\Delta E$ \\
\hline
\noalign{\smallskip}       
12 & $\mathrm{6d^2~^3P}$ & 0 & + &  5090   &  6151  &  -1061 &&      &      &&       &       &&       &        && 5641  & -551  \\
32 & $\mathrm{7s^2~^1S}$ & 0 & + &  11961  &  12428 &   -467 &&      &      && 15309 & -3348 &&  12459& -498   && 12909 & -948  \\
44 & $\mathrm{6d^2~^1S}$ & 0 & + &  18993  &  22008 &  -3015 &&      &      &&       &       &&       &        && 22547 & -3554 \\
48 & $\mathrm{5f^2~^3P}$ & 0 & + &  29300  &  29579 &   -279 &&      &      &&       &       &&       &        && 30429 & -1129 \\
63 & $\mathrm{5f^2~^1S}$ & 0 & + &  51162  &  55356 &  -4194 &&      &      &&       &       &&       &        && 55047 & -3885 \\     
13 & $\mathrm{6d7s~^3D}$ & 1 & + &  5524   &  6137  &   -613 &&      &      && 6947  & -1423 &&  6520 & -996   && 5998  & -474  \\
20 & $\mathrm{6d^2~^3P}$ & 1 & + &  7876   &  8905  &  -1029 &&      &      && 9771  & -1895 &&  8272 & -396   && 8088  & -212  \\
49 & $\mathrm{5f^2~^3P}$ & 1 & + &  30403  &  30636 &   -233 &&      &      &&       &       &&       &        && 31111 & -708  \\
58 & $\mathrm{5f7p~^3D}$ & 1 & + &  44603  &  44946 &   -343 &&      &      &&       &       &&       &        && 44848 & -245  \\     
2  & $\mathrm{6d^2~^3F}$ & 2 & + &  63     &  895   &   -832 &&      &      &&    0  &   63  &&    0  &   63   && 266   & -203  \\
10 & $\mathrm{6d^2~^1D}$ & 2 & + &  4676   &  5426  &   -750 &&      &      && 5451  & -775  &&  5003 & -327   && 4926  & -250  \\
17 & $\mathrm{6d7s~^3D}$ & 2 & + &  7176   &  7943  &   -767 &&      &      && 8617  & -1441 &&  8011 & -835   && 7576  & -400  \\
27 & $\mathrm{6d^2~^3P}$ & 2 & + &  10440  &  11417 &   -977 &&      &      && 12356 & -1916 &&  11087& -647   && 10608 & -168  \\
36 & $\mathrm{6d7s~^1D}$ & 2 & + &  16038  &  16438 &   -400 &&      &      && 19163 & -3125 &&  17628& -1590  && 16265 & -227  \\
38 & $\mathrm{5f^2~^3F}$ & 2 & + &  18864  &  18616 &    248 &&      &      &&       &       &&       &        && 18869 & -5    \\
46 & $\mathrm{5f^2~^1D}$ & 2 & + &  28233  &  28971 &   -738 &&      &      &&       &       &&       &        && 29121 & -888  \\
50 & $\mathrm{5f^2~^3P}$ & 2 & + &  32867  &  33488 &   -621 &&      &      &&       &       &&       &        && 33345 & -478  \\
52 & $\mathrm{5f7p~^3F}$ & 2 & + &  34996  &        &        &&      &      &&       &       &&       &        && 35475 & -479  \\
57 & $\mathrm{5f7p~^3D}$ & 2 & + &  43759  &        &        &&      &      &&       &       &&       &        && 44021 & -262  \\
62 & $\mathrm{5f7p~^1D}$ & 2 & + &  49806  &        &        &&      &      &&       &       &&       &        && 50225 & -419  \\     
7  & $\mathrm{6d^2~^3F}$ & 3 & + &  4056   &  4938  &   -882 &&  4023&   33 && 4278  & -222  &&  4478 & -422   && 4162  & -106  \\
25 & $\mathrm{6d7s~^3D}$ & 3 & + &  9954   &  10641 &   -687 &&  9204&  750 && 11465 & -1511 &&  11038& -1084  && 10371 & -417  \\
41 & $\mathrm{5f^2~^3F}$ & 3 & + &  20840  &  20378 &    462 && 19068& 1772 &&       &       &&       &        && 20357 & 483   \\
51 & $\mathrm{5f7p~^3G}$ & 3 & + &  33562  &  33715 &   -153 &&      &      &&       &       &&       &        && 33944 & -382  \\
53 & $\mathrm{5f7p~^3F}$ & 3 & + &  38432  &  38736 &   -304 &&      &      &&       &       &&       &        && 38784 & -352  \\
55 & $\mathrm{5f7p~^1F}$ & 3 & + &  42313  &  42544 &   -231 &&      &      &&       &       &&       &        && 42459 & -146  \\
60 & $\mathrm{5f7p~^3D}$ & 3 & + &  47471  &  47876 &   -405 &&      &      &&       &       &&       &        && 47647 & -176  \\     
16 & $\mathrm{6d^2~^3F}$ & 4 & + &  6538   &  7264  &   -726 &&  6795&  -257&& 7385  & -847  &&  7174 & -636   && 6629  & -91   \\
26 & $\mathrm{6d^2~^1G}$ & 4 & + &  10543  &  10822 &   -279 && 11051&  -508&& 13567 & -3024 &&  11742& -1199  && 10439 & 104   \\
34 & $\mathrm{5f^2~^3H}$ & 4 & + &  15149  &  14514 &    635 && 13358&  1791&&       &       &&       &        && 14636 & 513   \\
42 & $\mathrm{5f^2~^3F}$ & 4 & + &  21784  &  21782 &      2 && 20366&  1418&&       &       &&       &        && 21504 & 280   \\
45 & $\mathrm{5f^2~^1G}$ & 4 & + &  25972  &  27045 &  -1073 && 25269&   703&&       &       &&       &        && 25815 & 157   \\
54 & $\mathrm{5f7p~^3F}$ & 4 & + &  38581  &  38980 &   -399 &&      &      &&       &       &&       &        && 39062 & -481  \\
56 & $\mathrm{5f7p~^3G}$ & 4 & + &  43702  &  44034 &   -332 &&      &      &&       &       &&       &        && 44001 & -299  \\
59 & $\mathrm{5f7p~^1G}$ & 4 & + &  47261  &  47745 &   -484 &&      &      &&       &       &&       &        && 47581 & -320  \\     
37 & $\mathrm{5f^2~^3H}$ & 5 & + &  17887  &  17131 &    756 && 16068&  1819&&       &       &&       &        && 17202 & 685   \\
61 & $\mathrm{5f7p~^3G}$ & 5 & + &  47422  &  47781 &   -359 &&      &      &&       &       &&       &        && 47956 & -534  \\     
40 & $\mathrm{5f^2~^3H}$ & 6 & + &  20771  &  20123 &    648 &&      &      &&       &       &&       &        && 19972 & 799   \\
47 & $\mathrm{5f^2~^1I}$ & 6 & + &  28350  &  28635 &   -285 &&      &      &&       &       &&       &        && 29422 & -1072 \\
\hline
\end{tabular}                        
\end{table*}

\begin{table}
\setlength{\tabcolsep}{3.1pt}
\caption{RMS (in cm$^{-1}$) of our computed energy levels, 
of CI$+$all method \citet{Safronova}, 
of CI$+$MBPT method \citet{Dzuba},
and of SA-CASSCF and MS-CASPT2 methods \citet{ROY201225}  
 from the experiment \citet{Redman_2014}, according configurations.}
\label{summary_accuracy}
\begin{tabular}{lrrrrrrrrrrrr}
\hline\hline
\noalign{\smallskip}
Conf.   &CI$+$all   && CI$+$MBPT &SA-CASSCF&MS-CASPT2   &RCI\\  
\hline                         
$\mathrm{5f6d}$  &  637 &&  ... & 6160 & 1885 & 337\\
$\mathrm{5f7s}$  &  137 &&  ... & 8570 & 3416 &  94\\
\hline                         
$\mathrm{6d^2}$  & 1331 &&  329 & 1740 &  636 &253$^1$(1208)\\            
$\mathrm{6d7s}$  &  580 &&  750 & 2010 & 1162 &391\\
$\mathrm{7s^2}$  &  467 &&  ... & 3348 &  498 &948\\  
$\mathrm{5f^2}$  & 1288 && 1560 & ...  &  ... &686$^1$(1263)\\ 
$\mathrm{5f7p}$  & 347  &&   86 &  ... &  ... &361\\
\hline                        
all              &  891 && 1089 & 5113 & 1871 &436$^2$(796)\\
\hline
\end{tabular}      
\end{table}   

\section{Results}
\label{sec:results}

\subsection{Comparison of Energy Levels}
\label{E_comparison}

The accuracy of our computed energy levels (RCI) is evaluated by comparing
with experimental data by \citet{Redman_2014} and theoretical
data. \cite{Safronova} studied energy levels of $\mathrm{5f\{7s,8s,6d,7d\}}$, and $\mathrm{6d7p}$ odd-parity 
and of $\mathrm{6d^2}$, $\mathrm{5f^2}$, $\mathrm{7s^2}$, $\mathrm{5f\{7p,6f\}}$, and $\mathrm{6d7s}$ even-parity configurations using CI$+$all approach,  
while \cite{Dzuba} have presented only eleven energy levels 
of $\mathrm{6d^2}$, $\mathrm{6d7s}$, $\mathrm{5f^2}$, and $\mathrm{5f7p}$ configurations, using CI$+$MBPT approach. 
\cite{ROY201225} have calculated a part of energy levels of $\mathrm{5f\{6d,7s\}}$  
 odd-parity and of $\mathrm{6d^2}$, $\mathrm{7s^2}$, and $\mathrm{6d7s}$ even-parity using 
SA-CASSCF and MS-CASPT2 methods.  
Our calculated energy levels and the differences 
from the experimental values are given in Tables \ref{Energies_o} and \ref{Energies_e}.
Theoretical values calculated by other authors 
and differences from experimental data are also given in the same tables.

\begin{table*}
\caption{ASFs labels for present (RCI) work and composition (in percentages) in $LS$-coupling 
from \citet{Blaise} and \citet{Safronova} (CI$+$all) studies.}
\label{Labels_LS}
\begin{tabular}{rlllc}
\hline\hline
\noalign{\smallskip}
\multicolumn{1}{c}{\multirow{2}{0.25cm}{No.~~}} 
& \multicolumn{1}{c}{\multirow{2}{1.50cm}{RCI labels}}
&\multicolumn{2}{c}{Compositions}   
& \multicolumn{1}{c}{\multirow{2}{1.50cm}{CI$+$all labels}} \\ \cline{3-4}
& & \multicolumn{1}{c}{RCI} & \multicolumn{1}{c}{\cite{Blaise}} & \\
\noalign{\smallskip}
\hline
\noalign{\smallskip}       
31 & $\mathrm{5f6d~^{3}P_{0}^{\circ}}$ & 92 & 99 + 1  $\mathrm{6d7p~^3P^{\circ}}$ & ...\\
21 & $\mathrm{5f6d~^{3}D_{1}^{\circ}}$ & 81 +  6~$\mathrm{5f6d~^{1}P^{\circ}}$ +  5~$\mathrm{5f6d~^{3}P^{\circ}}$ & 83 + 8  $\mathrm{5f6d~^1P^{\circ}}$ & $\mathrm{^{3}D_{1}^{\circ}}$\\
30 & $\mathrm{5f6d~^{3}P_{1}^{\circ}}$ & 80 +  8~$\mathrm{5f6d~^{3}D^{\circ}}$ +  4~$\mathrm{5f6d~^{1}P^{\circ}}$ & 83 + 11 $\mathrm{5f6d~^3D^{\circ}}$ & $\mathrm{^{3}P_{1}^{\circ}}$\\
43 & $\mathrm{5f6d~^{1}P_{1}^{\circ}}$ & 81 +  6~$\mathrm{5f6d~^{3}P^{\circ}}$ +  3~$\mathrm{5f6d~^{3}D^{\circ}}$ & 86 + 7  $\mathrm{5f6d~^3P^{\circ}}$ & $\mathrm{^{1}P_{1}^{\circ}}$\\
3  & $\mathrm{5f6d~^{3}F_{2}^{\circ}}$ & 61 + 20~$\mathrm{5f6d~^{1}D^{\circ}}$ + 11~$\mathrm{5f7s~^{3}F^{\circ}}$ & 60 + 23 $\mathrm{5f6d~^1D^{\circ}}$ & $\mathrm{^{3}F_{2}^{\circ}}$\\
6  & $\mathrm{5f7s~^{3}F_{2}^{\circ}}$ & 76 + 13~$\mathrm{5f6d~^{1}D^{\circ}}$ +  2~$\mathrm{5f6d~^{3}F^{\circ}}$ & 80 + 14 $\mathrm{5f6d~^1D^{\circ}}$ & $\mathrm{^{3}F_{2}^{\circ}}$\\
15 & $\mathrm{5f6d~^{1}D_{2}^{\circ}}$ & 49 + 29~$\mathrm{5f6d~^{3}F^{\circ}}$ +  6~$\mathrm{5f7s~^{3}F^{\circ}}$ & 49 + 35 $\mathrm{5f6d~^3F^{\circ}}$ & $\mathrm{^{1}D_{2}^{\circ}}$\\
24 & $\mathrm{5f6d~^{3}D_{2}^{\circ}}$ & 83 +  6~$\mathrm{5f6d~^{3}P^{\circ}}$ +  2~$\mathrm{5f6d~^{1}D^{\circ}}$ & 88 + 8  $\mathrm{5f6d~^3P^{\circ}}$ & $\mathrm{^{3}D_{2}^{\circ}}$\\
33 & $\mathrm{5f6d~^{3}P_{2}^{\circ}}$ & 79 +  8~$\mathrm{5f6d~^{1}D^{\circ}}$ +  4~$\mathrm{5f6d~^{3}D^{\circ}}$ & 84 + 9  $\mathrm{5f6d~^1P^{\circ}}$ & $\mathrm{^{3}P_{2}^{\circ}}$\\
4  & $\mathrm{5f7s~^{3}F_{3}^{\circ}}$ & 40 + 32~$\mathrm{5f6d~^{3}F^{\circ}}$ + 16~$\mathrm{5f7s~^{1}F^{\circ}}$ & 47 + 24 $\mathrm{5f6d~^3F^{\circ}}$ & $\mathrm{^{3}F_{3}^{\circ}}$\\
9  & $\mathrm{5f6d~^{3}F_{3}^{\circ}}$ & 34 + 25~$\mathrm{5f6d~^{3}G^{\circ}}$ + 18~$\mathrm{5f7s~^{3}F^{\circ}}$ & 40 + 31 $\mathrm{5f6d~^3G^{\circ}}$ & $\mathrm{^{1}F_{3}^{\circ}}$\\
11 & $\mathrm{5f6d~^{3}G_{3}^{\circ}}$ & 54 + 22~$\mathrm{5f6d~^{3}F^{\circ}}$ +  9~$\mathrm{5f7s~^{1}F^{\circ}}$ & 52 + 28 $\mathrm{5f6d~^3F^{\circ}}$ & $\mathrm{^{3}G_{3}^{\circ}}$\\
18 & $\mathrm{5f7s~^{1}F_{3}^{\circ}}$ & 43 + 35~$\mathrm{5f7s~^{3}F^{\circ}}$ +  6~$\mathrm{5f6d~^{3}G^{\circ}}$ & 43 + 39 $\mathrm{5f7s~^3F^{\circ}}$ & $\mathrm{^{1}F_{3}^{\circ}}$\\
28 & $\mathrm{5f6d~^{3}D_{3}^{\circ}}$ & 68 + 16~$\mathrm{5f6d~^{1}F^{\circ}}$ +  4~$\mathrm{5f7s~^{1}F^{\circ}}$ & 72 + 19 $\mathrm{5f6d~^1F^{\circ}}$ & $\mathrm{^{3}D_{3}^{\circ}}$\\
35 & $\mathrm{5f6d~^{1}F_{3}^{\circ}}$ & 61 + 23~$\mathrm{5f6d~^{3}D^{\circ}}$ +  5~$\mathrm{5f7s~^{1}F^{\circ}}$ & 67 + 25 $\mathrm{5f6d~^3D^{\circ}}$ & $\mathrm{^{3}F_{3}^{\circ}}$\\
1  & $\mathrm{5f6d~^{3}H_{4}^{\circ}}$ & 56 + 34~$\mathrm{5f6d~^{1}G^{\circ}}$ +  3~$\mathrm{5f6d~^{3}F^{\circ}}$ & 64 + 32 $\mathrm{5f6d~^1G^{\circ}}$ & $\mathrm{^{3}H_{4}^{\circ}}$\\
5  & $\mathrm{5f6d~^{1}G_{4}^{\circ}}$ & 42 + 36~$\mathrm{5f6d~^{3}H^{\circ}}$ + 14~$\mathrm{5f6d~^{3}F^{\circ}}$ & 49 + 34 $\mathrm{5f6d~^3H^{\circ}}$ & $\mathrm{^{1}G_{4}^{\circ}}$\\
14 & $\mathrm{5f7s~^{3}F_{4}^{\circ}}$ & 44 + 34~$\mathrm{5f6d~^{3}F^{\circ}}$ + 11~$\mathrm{5f6d~^{1}G^{\circ}}$ & 60 + 26 $\mathrm{5f6d~^3F^{\circ}}$ & $\mathrm{^{3}F_{4}^{\circ}}$\\
19 & $\mathrm{5f6d~^{3}G_{4}^{\circ}}$ & 72 + 15~$\mathrm{5f7s~^{3}F^{\circ}}$ +  5~$\mathrm{5f6d~^{3}F^{\circ}}$ & 75 + 14 $\mathrm{5f7s~^3F^{\circ}}$ & $\mathrm{^{3}G_{4}^{\circ}}$\\
23 & $\mathrm{5f6d~^{3}F_{4}^{\circ}}$ & 37 + 32~$\mathrm{5f7s~^{3}F^{\circ}}$ + 18~$\mathrm{5f6d~^{3}G^{\circ}}$ & 47 + 23 $\mathrm{5f7s~^3F^{\circ}}$ & $\mathrm{^{3}F_{4}^{\circ}}$\\
8  & $\mathrm{5f6d~^{3}H_{5}^{\circ}}$ & 93 & 99 + 1  $\mathrm{5f6d~^3G^{\circ}}$ & $\mathrm{^{3}H_{5}^{\circ}}$\\
29 & $\mathrm{5f6d~^{3}G_{5}^{\circ}}$ & 90 +  3~$\mathrm{5f6d~^{1}H^{\circ}}$ & 96 & $\mathrm{^{3}G_{5}^{\circ}}$\\
39 & $\mathrm{5f6d~^{1}H_{5}^{\circ}}$ & 89 +  3~$\mathrm{5f6d~^{3}G^{\circ}}$ & 96 + 3  $\mathrm{5f6d~^3G^{\circ}}$ & $\mathrm{^{1}H_{5}^{\circ}}$\\
22 & $\mathrm{5f6d~^{3}H_{6}^{\circ}}$ & 93 &100 & $\mathrm{^{3}H_{6}^{\circ}}$\\
\noalign{\smallskip}
\hline \noalign{\smallskip}
12 &$\mathrm{6d^{2}~^{3}P_{0}}$ & 66       + 12~$\mathrm{6d^{2}~^{1}S}$ +  8~$\mathrm{5f^{2}~^{3}P}$ & 65 +  16 $\mathrm{6d^2~^1S}$  & ...\\                     
32 &$\mathrm{7s^{2}~^{1}S_{0}}$ & 49       + 26~$\mathrm{6d^{2}~^{1}S}$ + 11~$\mathrm{6d^{2}~^{3}P}$ & 59 +  20 $\mathrm{6d^2~^1S}$  & ...\\                     
44 &$\mathrm{6d^{2}~^{1}S_{0}}$          & {\bf 29} + 36~$\mathrm{7s^{2}~^{1}S}$ + 21~$\mathrm{5f^{2}~^{1}S}$                  & 40 +  35 $\mathrm{7s^2~^1S}$  & ...\\
48 &$\mathrm{5f^{2}~^{3}P_{0}}$          & 78       + 10~$\mathrm{6d^{2}~^{3}P}$                                               & 82 +  14 $\mathrm{6d^2~^3P}$  & ...\\
63 &$\mathrm{5f^{2}~^{1}S_{0}}$          & 59       + 24~$\mathrm{6d^{2}~^{1}S}$ +  2~$\mathrm{7s^{2}~^{1}S}$                  & 75 +  21 $\mathrm{6d^2~^1S}$  & ...\\
13 &$\mathrm{6d7s~^{3}D_{1}}  $          & 90 & 97 +   2 $\mathrm{5f7p~^3D}$  & $\mathrm{^{3}D_{1}}$\\    
20 &$\mathrm{6d^{2}~^{3}P_{1}}$          & 80       + 11~$\mathrm{5f^{2}~^{3}P}$                                               & 84 +  15 $\mathrm{5f^2~^3P}$  & $\mathrm{^{3}P_{1}}$\\    
49 &$\mathrm{5f^{2}~^{3}P_{1}}$          & 79       + 11~$\mathrm{6d^{2}~^{3}P}$                                               & 84 +  15 $\mathrm{5f^2~^3P}$  & $\mathrm{^{3}P_{1}}$\\    
58 &$\mathrm{5f7p~^{3}D_{1}}$            & 92 & 97                            & $\mathrm{^{3}D_{1}}$\\
2  &$\mathrm{6d^{2}~^{3}F_{2}}$          & 65       + 17~$\mathrm{6d^{2}~^{1}D}$ +  4~$\mathrm{5f^{2}~^{3}F}$                  & 65 +  21 $\mathrm{6d^2~^1D}$  & $\mathrm{^{3}F_{2}}$\\    
10 &$\mathrm{6d^{2}~^{1}D_{2}}$          & 29       + 18~$\mathrm{6d7s~^{1}D}$   + 17~$\mathrm{6d^{2}~^{3}F}$                  & 32 +  22 $\mathrm{6d^2~^3F}$  & $\mathrm{^{1}D_{2}}$\\    
17 &$\mathrm{6d7s~^{3}D_{2}}  $          & 64       + 11~$\mathrm{6d^{2}~^{3}P}$ +  9~$\mathrm{6d^{2}~^{1}D}$                  & 75 +   7 $\mathrm{6d^2~^3P}$  & $\mathrm{^{3}D_{2}}$\\    
27 &$\mathrm{6d^{2}~^{3}P_{2}}$          & 58       +  9~$\mathrm{5f^{2}~^{3}P}$ +  9~$\mathrm{6d7s~^{1}D}$                    & 64 +  13 $\mathrm{5f^2~^3P}$  & $\mathrm{^{3}P_{2}}$\\    
36 &$\mathrm{6d7s~^{1}D_{2}}  $          & 35       + 20~$\mathrm{5f^{2}~^{1}D}$ + 16~$\mathrm{5f^{2}~^{3}F}$                  & 51 +  19 $\mathrm{5f^2~^1D}$  & $\mathrm{^{1}D_{2}}$\\    
38 &$\mathrm{5f^{2}~^{3}F_{2}}$          & 66       + 14~$\mathrm{6d7s~^{1}D}$   +  5~$\mathrm{6d^{2}~^{1}D}$                  & 82 +   7 $\mathrm{6d7s~^1D}$  & $\mathrm{^{3}F_{2}}$\\    
46 &$\mathrm{5f^{2}~^{1}D_{2}}$          & 44       + 23~$\mathrm{5f^{2}~^{3}P}$ + 11~$\mathrm{6d^{2}~^{1}D}$                  & 50 +  23 $\mathrm{5f^2~^3P}$  & $\mathrm{^{3}D_{2}}$\\    
50 &$\mathrm{5f^{2}~^{3}P_{2}}$          & 55       + 19~$\mathrm{5f^{2}~^{1}D}$ +  9~$\mathrm{6d^{2}~^{3}P}$                  & 59 +  21 $\mathrm{5f^2~^1D}$  & $\mathrm{^{3}P_{2}}$\\    
52 &$\mathrm{5f7p~^{3}F_{2}}  $          & 61       + 15~$\mathrm{5f7p~^{1}D}$   + 13~$\mathrm{5f7p~^{3}D}$                    & 63                            & ...\\
57 &$\mathrm{5f7p~^{3}D_{2}}  $          & 51       + 28~$\mathrm{5f7p~^{3}F}$   + 13~$\mathrm{5f7p~^{1}D}$                    & 51                            & ...\\
62 &$\mathrm{5f7p~^{1}D_{2}}  $          & 59       + 28~$\mathrm{5f7p~^{3}D}$   +  3~$\mathrm{5f7p~^{3}F}$                    & 61                            & ...\\
7  &$\mathrm{6d^{2}~^{3}F_{3}}$          & 85       +  6~$\mathrm{5f^{2}~^{3}F}$                                               & 91 +   8 $\mathrm{5f^2~^3F}$  & $\mathrm{^{3}F_{3}}$\\    
25 &$\mathrm{6d7s~^{3}D_{3}}  $          & 89 & 96 +   3 $\mathrm{5f7p~^3D}$  & $\mathrm{^{3}D_{3}}$\\    
41 &$\mathrm{5f^{2}~^{3}F_{3}}$          & 85       +  6~$\mathrm{6d^{2}~^{3}F}$                                               & 91 +   8 $\mathrm{6d^2~^3F}$  & $\mathrm{^{3}F_{3}}$\\    
51 &$\mathrm{5f7p~^{3}G_{3}}  $          & 62       + 23~$\mathrm{5f7p~^{1}F}$   +  7~$\mathrm{5f7p~^{3}F}$                    & 68 $\mathrm{5f7p~^3G}$        & $\mathrm{^{3}G_{3}}$\\    
53 &$\mathrm{5f7p~^{3}F_{3}}  $          & 41       + 32~$\mathrm{5f7p~^{3}D}$   + 18~$\mathrm{5f7p~^{1}F}$                    & 42 $\mathrm{5f7p~^3F}$        & $\mathrm{^{3}F_{3}}$\\    
55 &$\mathrm{5f7p~^{1}F_{3}}  $          & {\bf 28} + 34~$\mathrm{5f7p~^{3}F}$   + 29~$\mathrm{5f7p~^{3}G}$                    & 37 $\mathrm{5f7p~^3F}$        & $\mathrm{^{1}F_{3}}$\\    
60 &$\mathrm{5f7p~^{3}D_{3}}  $          & 55       + 25~$\mathrm{5f7p~^{1}F}$   + 11~$\mathrm{5f7p~^{3}F}$                    & 55 $\mathrm{5f7p~^3D}$        & $\mathrm{^{3}F_{3}}$\\
16 &$\mathrm{6d^{2}~^{3}F_{4}}$          & 62       + 17~$\mathrm{6d^{2}~^{1}G}$ +  6~$\mathrm{5f^{2}~^{3}F}$                  & 69 +  17 $\mathrm{6d^2~^1G}$  & $\mathrm{^{3}F_{4}}$\\    
26 &$\mathrm{6d^{2}~^{1}G_{4}}$          & 43       + 22~$\mathrm{6d^{2}~^{3}F}$ + 20~$\mathrm{5f^{2}~^{1}G}$                  & 59 +  20 $\mathrm{6d^2~^3F}$  & $\mathrm{^{3}G_{4}}$\\    
34 &$\mathrm{5f^{2}~^{3}H_{4}}$          & 84       +  6~$\mathrm{6d^{2}~^{1}G}$                                               & 93 +   4 $\mathrm{6d^2~^1G}$  & $\mathrm{^{3}H_{4}}$\\    
42 &$\mathrm{5f^{2}~^{3}F_{4}}$          & 68       + 12~$\mathrm{5f^{2}~^{1}G}$ +  6~$\mathrm{6d^{2}~^{1}G}$                  & 63 +  22 $\mathrm{5f^2~^1G}$  & $\mathrm{^{3}F_{4}}$\\    
45 &$\mathrm{5f^{2}~^{1}G_{4}}$          & 54       + 18~$\mathrm{6d^{2}~^{1}G}$ + 16~$\mathrm{5f^{2}~^{3}F}$                  & 57 +  26 $\mathrm{5f^2~^3F}$  & $\mathrm{^{1}G_{4}}$\\    
54 &$\mathrm{5f7p~^{3}F_{4}}  $          & {\bf 31}       + 44~$\mathrm{5f7p~^{3}G}$   + 17~$\mathrm{5f7p~^{1}G}$                    & 46 $\mathrm{5f7p~^3G}$        & $\mathrm{^{3}G_{4}}$\\    
56 &$\mathrm{5f7p~^{3}G_{4}}  $          & 47       + 27~$\mathrm{5f7p~^{1}G}$   + 18~$\mathrm{5f7p~^{3}F}$                    & 51 $\mathrm{5f7p~^3G}$        & $\mathrm{^{3}G_{4}}$\\    
59 &$\mathrm{5f7p~^{1}G_{4}}  $          & 47       + 43~$\mathrm{5f7p~^{3}F}$                                                 & 51 $\mathrm{5f7p~^3F}$        & $\mathrm{^{3}F_{4}}$\\
37 &$\mathrm{5f^{2}~^{3}H_{5}}$          & 92 &100                            & $\mathrm{^{3}H_{5}}$\\    
61 &$\mathrm{5f7p~^{3}G_{5}}  $          & 93 &100                            & $\mathrm{^{3}G_{5}}$\\
40 &$\mathrm{5f^{2}~^{3}H_{6}}$          & 91 & 97 +   2 $\mathrm{5f^2~^1I}$  & $\mathrm{^{3}H_{6}}$\\    
47 &$\mathrm{5f^{2}~^{1}I_{6}}$          & 90 & 97 +   2 $\mathrm{5f^2~^3H}$  & $\mathrm{^{1}I_{6}}$\\
\hline
\end{tabular}
\end{table*}

The differences from the experiment are more significant than 1000 cm$^{-1}$ 
for eight levels of the CI$+$all method,
 and for four energy levels for the CI$+$MBPT method.  
Meanwhile, almost all levels of the MS-CASPT2 method 
differ from the experiment by more than a thousand cm$^{-1}$. 
The differences are even worse for the levels from the SA-CASSCF method. 
It should also be noted that the ground 
state was $\mathrm{6d^2~^3F_2}$ in both calculations by \cite{ROY201225}, and this 
 contradicts the most recent data from \cite{Redman_2014}. 
In our study, only five energies differ from the experimental data of 
\cite{Redman_2014} of more than 1000 cm$^{-1}$. The ground state is also 
consistent with the experimental findings. 

A summary of RMS from the experimental data is given for each configuration, and all levels are compared in Table \ref{summary_accuracy} for all authors.
The superscript $^1$ in Table \ref{summary_accuracy} 
indicate that the RMS is calculated excluding one energy level, which is more distant from experimental values than the others in that configuration.
The energy value of our level $\mathrm{6d^2~^1S_0}$ is 22547 cm$^{-1}$ 
and this value is larger than the \cite{Redman_2014} value by 3554 cm$^{-1}$ 
and also for level $\mathrm{5f^2~^1S_0}$ our calculated
 energy of 55047 is 3885 cm$^{-1}$ higher. RMS calculated with all levels of configuration are given in parentheses.

The RMS deviation of all compared levels is given in the last line ("all") of Table \ref{summary_accuracy}.  
The RMS deviation of all our levels from experimental data is 436$^2$(796)  
while the deviation of levels 
from CI$+$all and CI$+$MBPT method are larger   
(891 and 1089).
The most significant deviation from experiments 
is visible for the MS-CASPT2 method (1871) results. The superscript $^2$ in the table indicates that the RMS is calculated excluding two energy levels mentioned in the above paragraph. 

The labels of atomic state functions and composition in $LS-$, $JJ-$couplings of our computations and \citet{Blaise} 
and \citet{Safronova} are given in Tables \ref{Labels_LS} and \ref{Labels_JJ}. 
It is important to note that, in a general sense, a $jj$-coupling is not 
identical to a $JJ$-coupling, and that the notation of the terms is also different \citep{Martin,GAIGALAS_couling}. 
However, in the present case (5f7p - two orbitals with one electron each) 
the angular momenta are the same in both couplings, therefore the $jj$-coupling corresponds to the $JJ$-coupling \citep{GAIGALAS_couling}. 
This allows us to use \citet{Blaise} $JJ$-coupling identification instead of 
$jj$-coupling to analyze the purity of ASF in relativistic scheme calculation.

Labels of the atomic state function (configuration and term) 
for odd configurations are identical to those presented in \citet{Blaise} work, even in instances where the CSF 
mixing is more pronounced (see Table \ref{Labels_LS}). 
 The ASFs of even parity are observed to be more mixed than those of odd parity. 
 Contributions of configuration state functions are marked in bold for the states in which the labels are not assigned with the
most significant contribution to the composition.
Only the first components of the composition in the $LS-$coupling for ASFs of the $\mathrm{5f7p}$ configuration are provided by \citet{Blaise}, 
who also based their findings on these terms. 
However, it should be noted that the identification of ASFs given by \citet{Blaise} is not unique. 
In the meantime, we have transformed the ASFs from $jj-$ to the $LS-$coupling with a 
unique representation, as proposed by \cite{unic}. 
 The ASFs for this configuration are purer in the $JJ-$coupling,
 as demonstrated by \citet{Blaise}. Our computations indicate that the atomic state functions are also less mixed in the $JJ-$coupling than in the $LS-$coupling. 
However, they remain more mixed than in the \citet{Blaise} 
study (see Table \ref{Labels_JJ}).
The ASF labels are provided in the $LS-$coupling from the CI$+$all
 method without the composition. Sometimes, the terms 
 in the $LS-$coupling from the CI$+$all 
method differ from our results and those presented in \citet{Blaise}.     

\begin{table}                  
\caption{Atomic state function labels 
in $LS-$coupling and in $JJ-$coupling 
and the compositions in percentages
 from present computation (RCI) and given by \citet{Blaise} (Ref.).}
\label{Labels_JJ}            
\begin{tabular}{lllrr}  
\hline\hline      
\noalign{\smallskip}
\multicolumn{1}{c}{\multirow{2}{0.25cm}{No.~~}} &
\multicolumn{1}{c}{\multirow{2}{1.50cm}{RCI label $LS$}} 
& \multicolumn{1}{c}{\multirow{2}{1.50cm}{RCI label $JJ$}}
&\multicolumn{2}{c}{$JJ$ Composition in \%} \\
\cline{4-5}
& & &\multicolumn{1}{r}{RCI} & \multicolumn{1}{r}{Ref.} \\
\hline
\noalign{\smallskip}       
58 & $5f\,7p~^{3}D_{1}$ & $5f\,7p~(5/2,3/2)_{1}$ &   93 & 100  \\
52 & $5f\,7p~^{3}F_{2}$ & $5f\,7p~(5/2,1/2)_{2}$ &   83 &  96  \\
57 & $5f\,7p~^{3}D_{2}$ & $5f\,7p~(5/2,3/2)_{2}$ &   88 &  96  \\
62 & $5f\,7p~^{1}D_{2}$ & $5f\,7p~(7/2,3/2)_{2}$ &   89 &  99  \\
51 & $5f\,7p~^{3}G_{3}$ & $5f\,7p~(5/2,1/2)_{3}$ &   92 &  99  \\
53 & $5f\,7p~^{3}F_{3}$ & $5f\,7p~(7/2,1/2)_{3}$ &   88 &  96  \\
55 & $5f\,7p~^{1}F_{3}$ & $5f\,7p~(5/2,3/2)_{3}$ &   91 &  98  \\
60 & $5f\,7p~^{3}D_{3}$ & $5f\,7p~(7/2,3/2)_{3}$ &   88 &  96  \\
54 & $5f\,7p~^{3}F_{4}$ & $5f\,7p~(7/2,1/2)_{4}$ &   89 &  96  \\
56 & $5f\,7p~^{3}G_{4}$ & $5f\,7p~(5/2,3/2)_{4}$ &   86 &  88  \\
59 & $5f\,7p~^{1}G_{4}$ & $5f\,7p~(7/2,3/2)_{4}$ &   84 &  87  \\
61 & $5f\,7p~^{3}G_{5}$ & $5f\,7p~(7/2,3/2)_{5}$   & 93 & 100  \\
\hline
\end{tabular}
\end{table}

\begin{figure}
\centering
\includegraphics[width=\hsize]{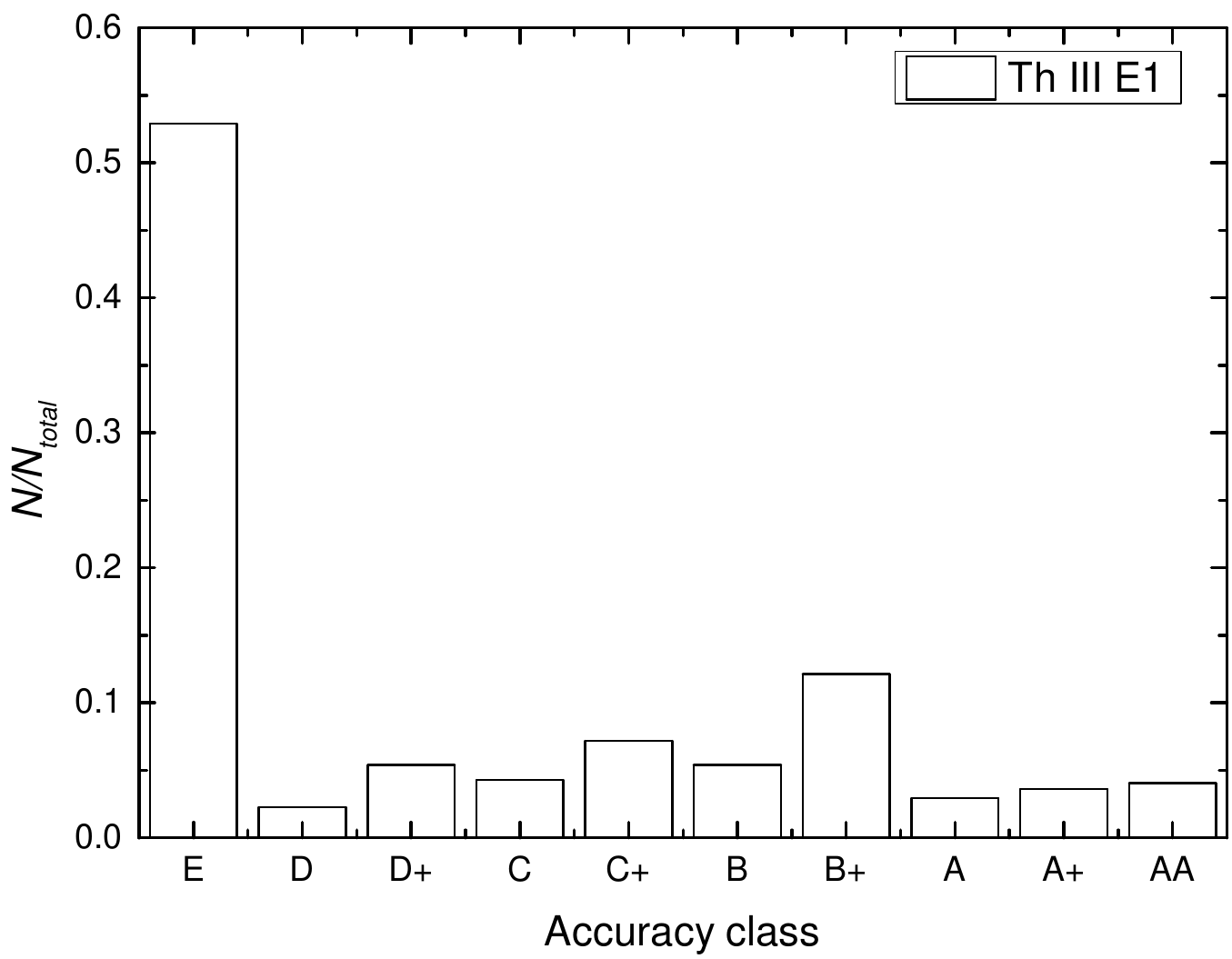}
\caption{\label{Acc_class}Distribution of E1 transitions over accuracy classes for Th\,{\sc iii}. 
The ratio of the number of transitions in a given accuracy class ($N$)
to the total number ($N_{total}$) of transitions in the y-axis is given.}
\end{figure}

\section{Evaluation of the line strengths}
\label{S_evealuattion}
In this work we  have calculated 455 E1-type transitions between the energy levels of the configurations mentioned above. 
QQE method~\citep{Pavel_Ce,Pavel_Pr,Kitoviene_Ge} was used to evaluate the uncertainty of the calculated line strengths.
Using the QQE method, all these transitions were distributed according to the accuracy classes in Fig. \ref{Acc_class}. 
As can be seen, a majority of the transitions are in the E class. 
According to the QQE method, 46 percent fall into the D and better accuracy class.  

\begin{figure}
\centering
\includegraphics[width=\hsize]{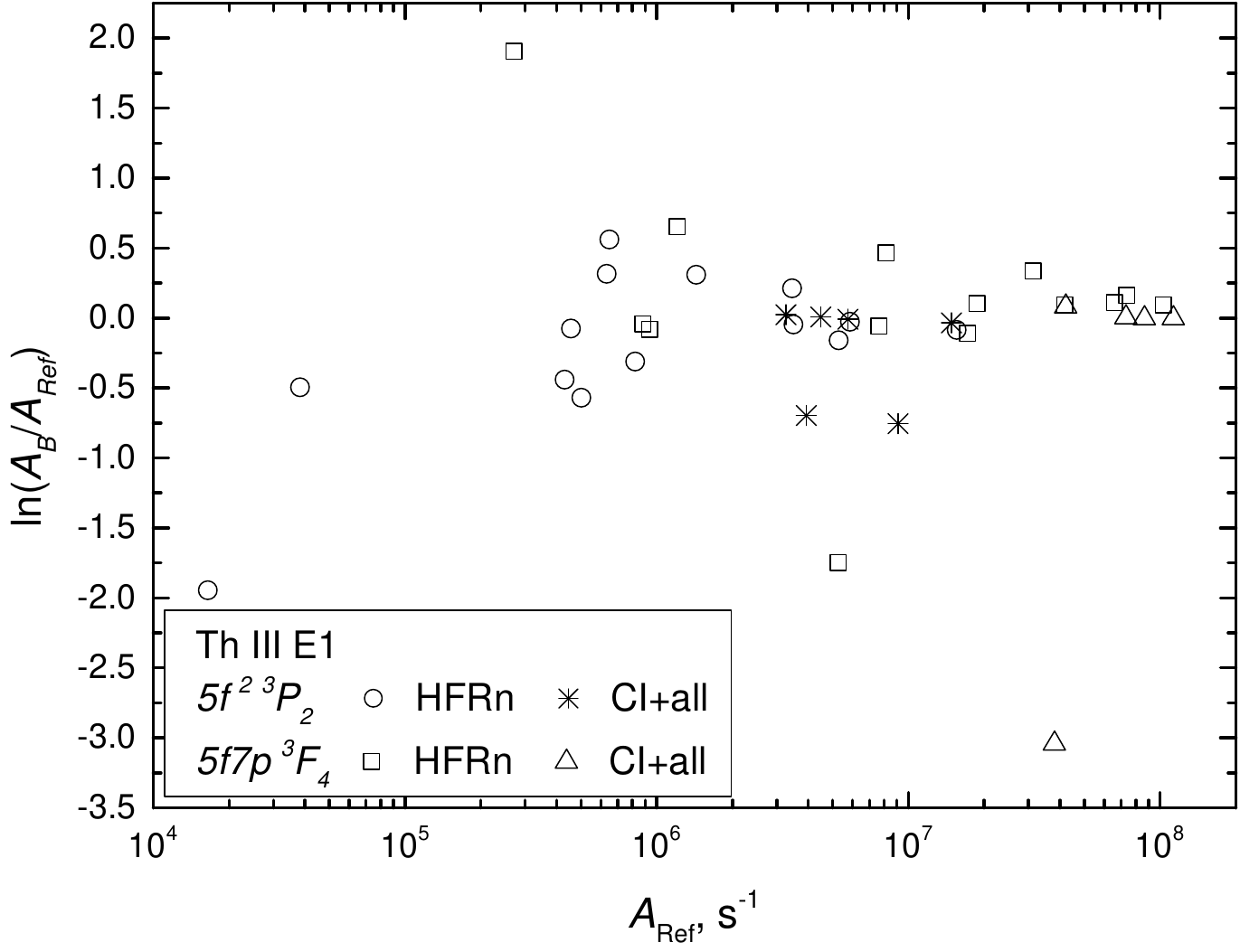}
\caption{\label{A_comparison_Safronova_Biemont} Comparison of our transition probabilities ($A_B$) depopulating energy levels:
 $\mathrm{5f^2~^3P_2}$ and $\mathrm{5f\,7p~^{3}F_{4}}$
for which the radiative lifetimes have been measured by \citet{Biemont}
with probabilities normalized to these lifetimes ($A_{HFRn}$) and 
with probabilities from CI$+$all method \citet{Safronova}.}
\end{figure}

Furthermore, our calculated transition data have been compared with the results of studies carried out by other authors. 
Initially, a comparison was made between the transition probabilities that originated from two states, specifically $\mathrm{5f^2~^3P_2}$ and $\mathrm{5f7p~^{3}F_{4}}$.  
The HFR method was employed by \citet{Biemont} to compute the transition probabilities 
for these states and then normalized according 
to their experimental lifetimes. These transition probabilities are compared in Fig. \ref{A_comparison_Safronova_Biemont}. 
As we can see, despite a couple of transitions from state $\mathrm{5f7p~^{3}F_{4}}$, 
the coincidence of probabilities is perfect.
Our transition probabilities were also compared with 
probabilities from CI$+$all method \citet{Safronova} 
for the same states.  Similarly to the comparison with \citet{Biemont} results, only one transition disagrees.     

\begin{figure}
\centering
\includegraphics[width=\hsize]{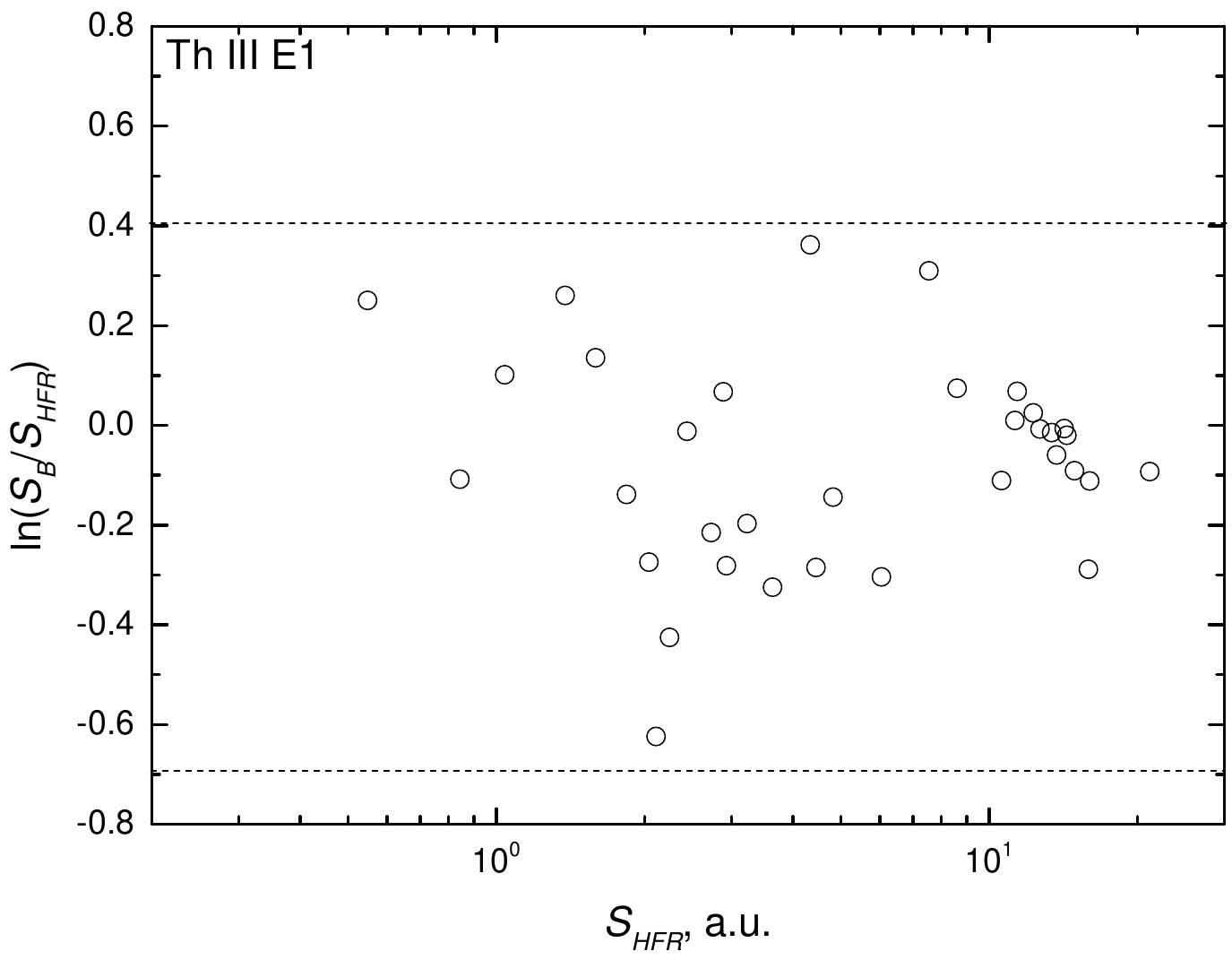}
\caption{\label{S_comparison_Biemont} Comparison of our 
 line strengths ($S_B$) with computed using HFR method ($S_{HFR}$) 
by \citet{Biemont} for the strongest transitions.}
\end{figure}

\begin{figure}
\centering
\includegraphics[width=\hsize]{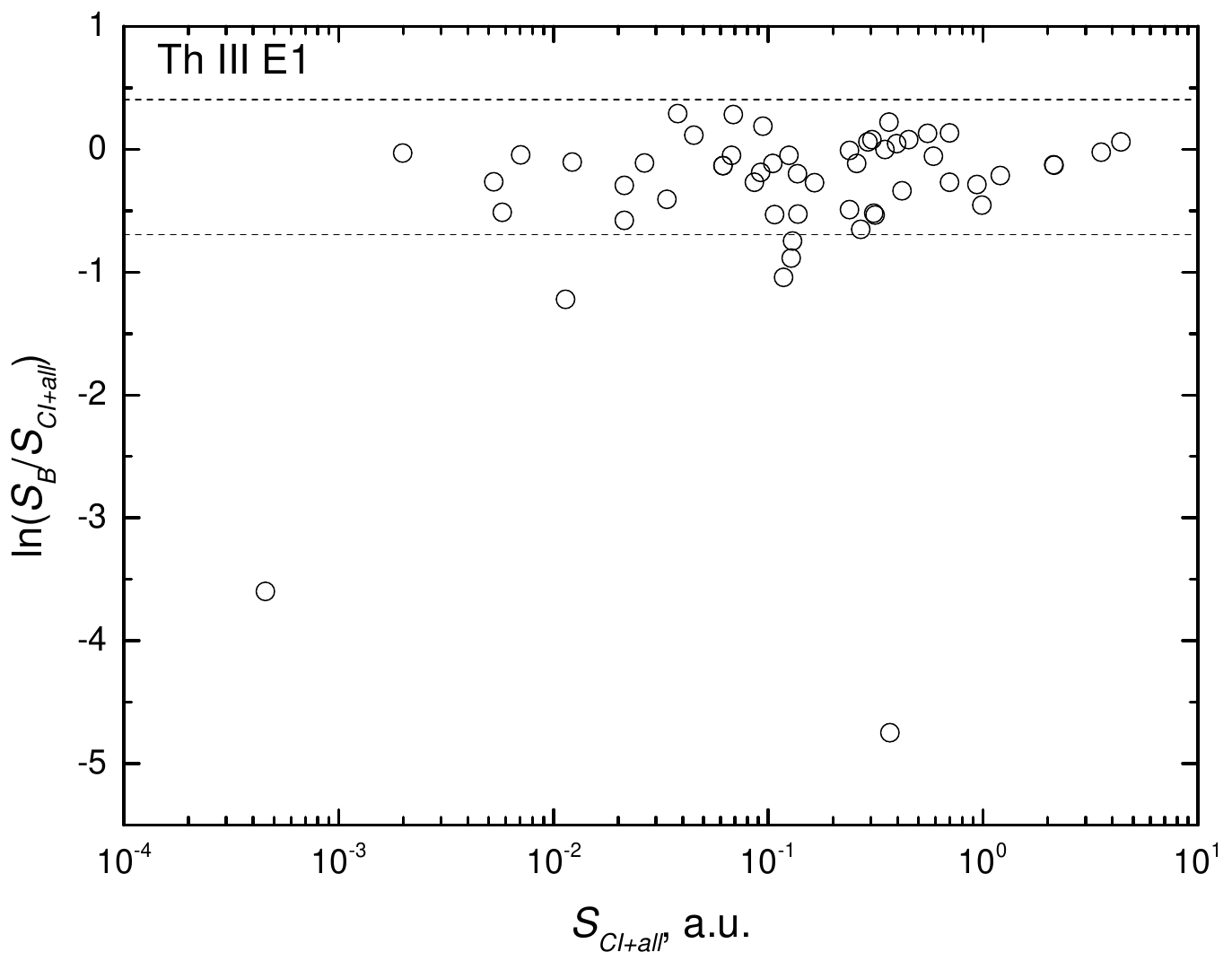}
\caption{\label{S_comparison_Safronova} Comparison of our 
 line strengths ($S_B$) with computed using CI$+$all method ($S_{CI+all}$) 
by \citet{Safronova}.}
\end{figure}

A comparison of the lines strengths of \citet{Biemont} that were computed with HFR 
(see Table 5 in reference to \citet{Biemont}) 
 was also performed  
in Fig. \ref{S_comparison_Biemont}. As can be seen, 
the agreement between the line strengths is satisfactory. 
Line strengths were also compared with data by
 \citet{Safronova} in Fig. \ref{S_comparison_Safronova}. 
Two transitions disagree greatly with our computed one, while others agree well.  

\begin{figure}
\centering
\includegraphics[width=\hsize]{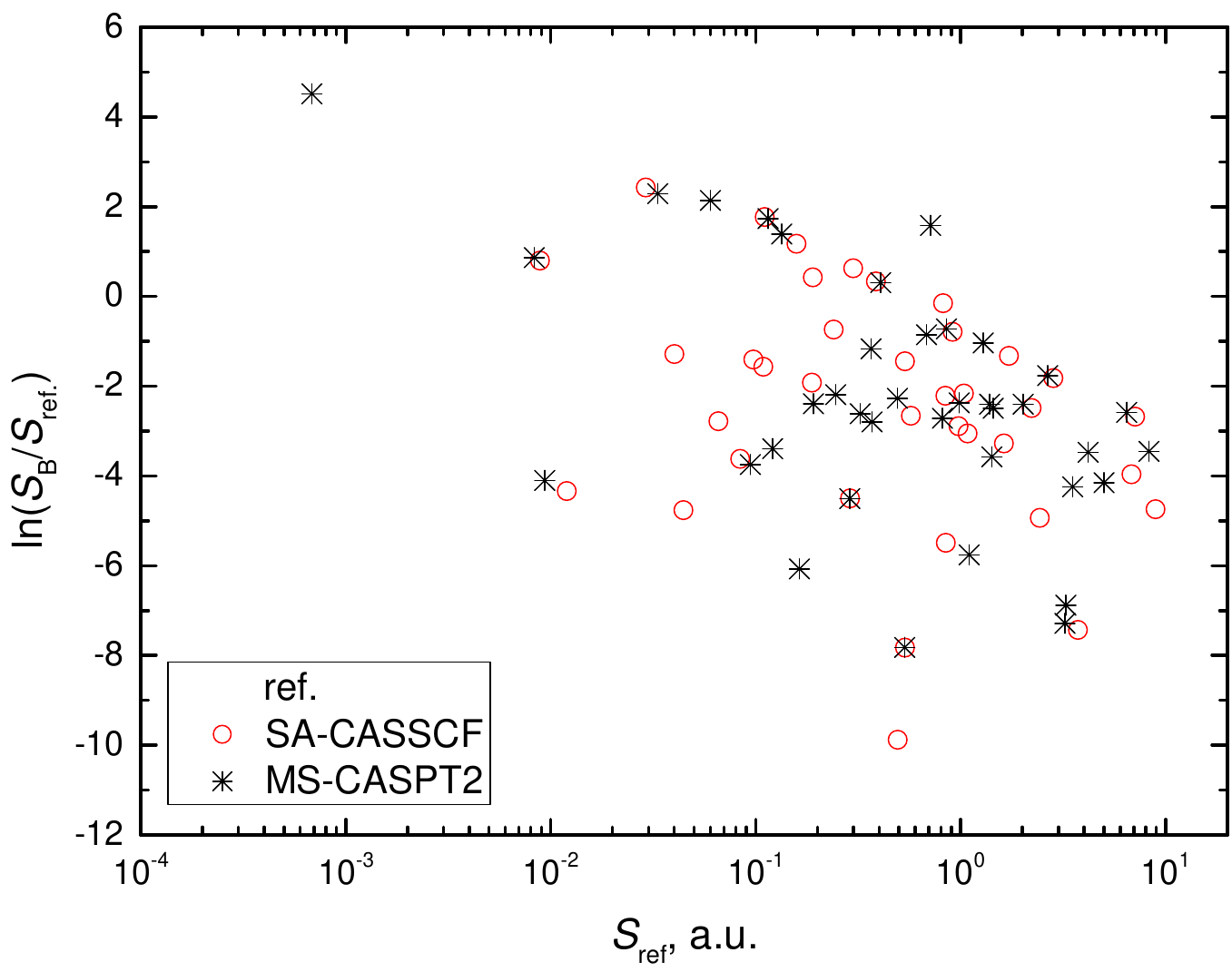}
\caption{\label{S_comparison_Roy} Comparison of our 
line strengths ($S_B$) with $S_{SA-CASSCF}$ and $S_{MS-CASPT2}$ computed by \citet{ROY201225}.}
\end{figure}

Line strengths are also compared 
with data from $S_{SA-CASSCF}$ and $S_{MS-CASPT2}$ methods computed 
by \citet{ROY201225} in Fig. \ref{S_comparison_Roy}. The line strengths are 
more scattered than the data from other methods.

 Transition data are given in Table \ref{Transition} (the complete table can be found in the supplementary material).  
 Level indexes (of lower and upper level $No_l$ and $No_u$ respectively are the same as in the Tables \ref{Energies_e} and \ref{Energies_o}), wavelengths, 
transition type, branching fractions $\left( BF=A_{ji}/{\sum}_{k=1}^{j-1}A_{jk}\right)$ \citep{Wang_2018}, rates, weighted oscillator strengths, line strengths 
(the last three are given in the Babushkin and Coulomb gauges) and accuracy classes for line strengths are provided in this table.

\subsection{Lifetimes}
Our computed lifetimes ($\tau$) are compared with the measured 
values ($\tau_{LIF}$) by \citet{Biemont} 
in Table \ref{lifetimes}. 
In addition to our {\em ab-initio} RCI$_B$ and RCI$_C$ calculations, we have computed the lifetimes (RCI$_{Bex}$) using experimental wavelengths for transition probabilities.
Lifetimes computed using HFR by \citet{Biemont} are given in the same table. 
 \cite{Safronova} have computed transition probabilities using experimental wavelengths and theoretical line strengths to evaluate the lifetimes ($\tau_{CI+all}$).   
These lifetimes are compared with experimental ones in Table \ref{lifetimes}. The lifetime in the Babushkin gauge (RCI$_B$) 
of the first state is closer to the experimental value than the values of other methods. For the second state, the value in the Babushkin gauge is smaller than values from other methods.

\begin{table}
\setlength{\tabcolsep}{5.3pt}
\caption{Our computed lifetimes (in ns) in Babushkin and Coulomb gauges (RCI$_{B,C}$)
compared with CI$+$all calculations \citet{Safronova} and 
LIF experimental and HFR data by \citet{Biemont}.}
\label{lifetimes}
\begin{tabular}{lrrrrrrrrrrrrrr}
\hline\hline
\noalign{\smallskip}
\multicolumn{1}{c}{Label} & \multicolumn{1}{c}{$CI+all$} & \multicolumn{1}{c}{HFR}  & \multicolumn{1}{c}{LIF} & \multicolumn{1}{c}{RCI$_B$} & \multicolumn{1}{c}{RCI$_C$} & \multicolumn{1}{c}{RCI$_{Bex}$} \\
\hline
\noalign{\smallskip}
$\mathrm{5f^2~^3P_2}$    &  21.2     & 23.7 & 25.8 $\pm$ 1.5 &  26.48     &39.70      &27.52\\
$\mathrm{5f7p~^3F_4}$    &  2.41     &  2.6 & 2.7  $\pm$ 0.2 &  2.35      &2.27       &2.46\\
\hline
\end{tabular}      
\end{table}   

\begin{table*}
\caption{Transition wavelengths $\lambda$ (in \AA), branching fractions BF, transition rates $A$ (in s$^{-1}$), weighted oscillator strengths $gf$, and line strengths $S$ (in a.u.) in the Babushkin (B) and Coulomb gauges (C), accuracy classes (Acc. class) for E1 transitions of the Th\,{\sc iii} [The full table is available online as supplementary data]. }
\label{Transition}
\begin{tabular}{ccrllllllllc}
\hline\hline
\noalign{\smallskip}       
 \multicolumn{1}{c}{No$_l$} &  \multicolumn{1}{c}{No$_u$} & \multicolumn{1}{c}{$\lambda$ (\AA)} & \multicolumn{1}{c}{BF} & \multicolumn{1}{c}{$A_B$ (s$^{-1}$)} & \multicolumn{1}{c}{$gf_B$} &  \multicolumn{1}{c}{$S_B$ (a.u.)} & \multicolumn{1}{c}{$A_C$ (s$^{-1}$)}  & \multicolumn{1}{c}{$gf_C$} & \multicolumn{1}{c}{$S_C$ (a.u.)} & \multicolumn{1}{c}{Acc. class}   \\
\hline
\noalign{\smallskip}
1 &  7   & 24 025.06~& 2.004E-04 & 2.605E-01 & 1.578E-07 & 1.248E-05 & 6.533E+02 & 3.957E-04 & 3.130E-02 & E~~\, \\
1 & 16~~ & 15 084.46~& 4.439E-01 & 3.309E+03 & 1.016E-03 & 5.045E-02 & 6.994E+04 & 2.147E-02 & 1.066E+00 & E~~\, \\
1 & 25~~ & 9 642.616 & 1.300E-01 & 7.661E+03 & 7.475E-04 & 2.373E-02 & 8.834E+03 & 8.620E-04 & 2.736E-02 & C+ \\
1 & 26~~ & 9 579.755 & 3.207E-01 & 1.572E+04 & 1.947E-03 & 6.139E-02 & 2.644E+04 & 3.274E-03 & 1.032E-01 & E~~\, \\
1 & 34~~ & 6 832.501 & 4.274E-01 & 1.094E+06 & 6.891E-02 & 1.550E+00 & 3.438E+05 & 2.166E-02 & 4.872E-01 & E~~\,  \\
1 & 37~~ & 5 813.180 & 1.083E-04 & 2.585E+02 & 1.441E-05 & 2.757E-04 & 5.341E+03 & 2.977E-04 & 5.697E-03 & E~~\,  \\
1 & 41~~ & 4 912.219 & 3.698E-04 & 3.052E+03 & 7.728E-05 & 1.250E-03 & 1.063E+01 & 2.691E-07 & 4.353E-06 & E~~\,  \\
1 & 42~~ & 4 650.371 & 1.774E-01 & 1.695E+06 & 4.946E-02 & 7.573E-01 & 7.584E+05 & 2.213E-02 & 3.388E-01 & E~~\,  \\
1 & 45~~ & 3 873.652 & 2.034E-01 & 2.731E+06 & 5.529E-02 & 7.051E-01 & 1.278E+06 & 2.588E-02 & 3.300E-01 & E~~\,  \\
1 & 51~~ & 2 946.022 & 3.408E-01 & 1.425E+08 & 1.298E+00 & 1.259E+01 & 1.429E+08 & 1.301E+00 & 1.262E+01 & AA \\
\hline
\end{tabular}
\end{table*}

\section{Applications to kilonova spectra}

\begin{figure}
\centering
\includegraphics[width=\hsize]{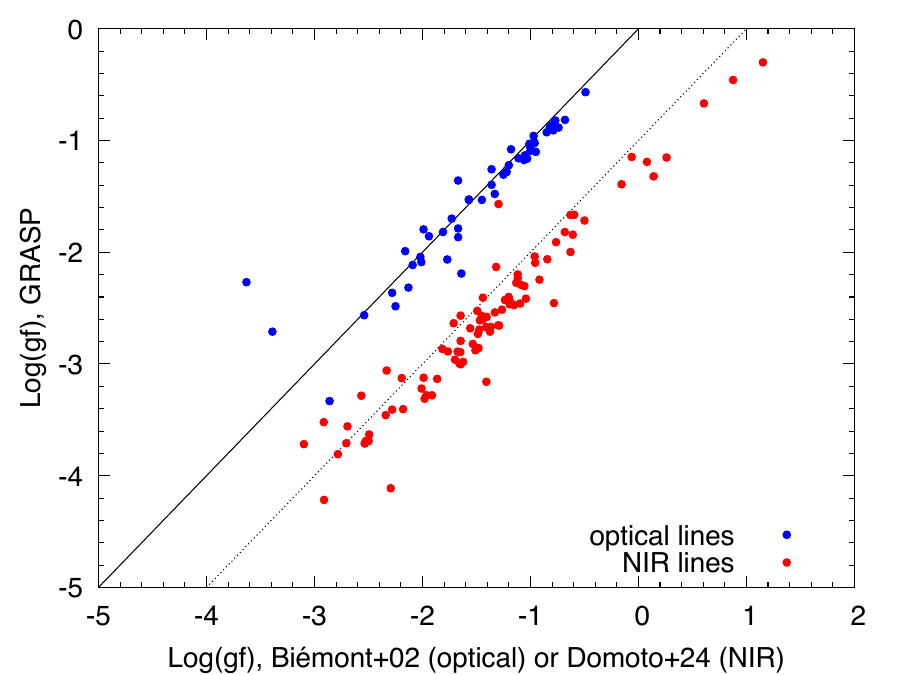}
\caption{\label{gfvs} Comparison of our 
log $gf$ values from the Babushkin gauge with those estimated by \citet{domoto24} for the lines listed in the NIST. Solid and dashed lines correspond to perfect agreement and deviation by a factor of 10, respectively.}
\end{figure}

In this section, we apply our calculated transition data to kilonova spectra.
To discuss the spectral features, we first calibrate the wavelengths of our transition data by cross-matching the transitions based on the $LS$ terms.
By this calibration, transition data used in this section have spectroscopically accurate wavelengths.
When the wavelengths of the transitions are corrected, we also correct the line strength according to the energy difference ($S \propto \Delta E^{-3}$).
However, this correction is insignificant and does not significantly impact the results as the accuracy in the energy is good enough.

Fig. \ref{gfvs} shows a comparison between the weighted oscillator strengths (log $gf$) from the Babushkin gauge and those estimated by \citet{domoto24}.
The estimates by \citet{domoto24} are based on the emission line ratios measured in the thorium-argon hollow cathode lamp spectrum \citet{engleman03} and calculated transition probabilities in optical wavelengths by \citet{Biemont_2002_MNRAS}.
As shown in the figure, our calculated $gf$ values in the NIR wavelengths are generally smaller than those estimated by \citet{domoto24} by a factor of about 10.

We perform radiative transfer simulations using our transition data to study the impacts on the spectral features.
We adopt the same setting of the calculations with \citet{domoto24}. 
Namely, we assume a one-dimensional, spherical ejecta with a power-law density structure ($\rho \propto v^{-3}$).
We adopt a spatially homogeneous abundance distribution.
We use their "Light" model for the abundance pattern, similar to metal-poor stars with weak $r$-process signature. In this abundance model, the mass fractions of the elements $Z>50$ are reduced by a factor of about 100 concerning the solar abundance ratio).
The mass fraction is $1.0 \times 10^{-5}$.
For this model, we perform radiative transfer simulations using a time-dependent, wavelength-dependent radiative transfer code \citep{tanaka13,Kawaguchi18}.
We use the same line list with \citet{domoto24} but adopt our transition data only for Th\,{\sc iii} lines.
We refer the reader to \citet{domoto24} for more details.

\begin{figure}
\centering
\includegraphics[width=\hsize]{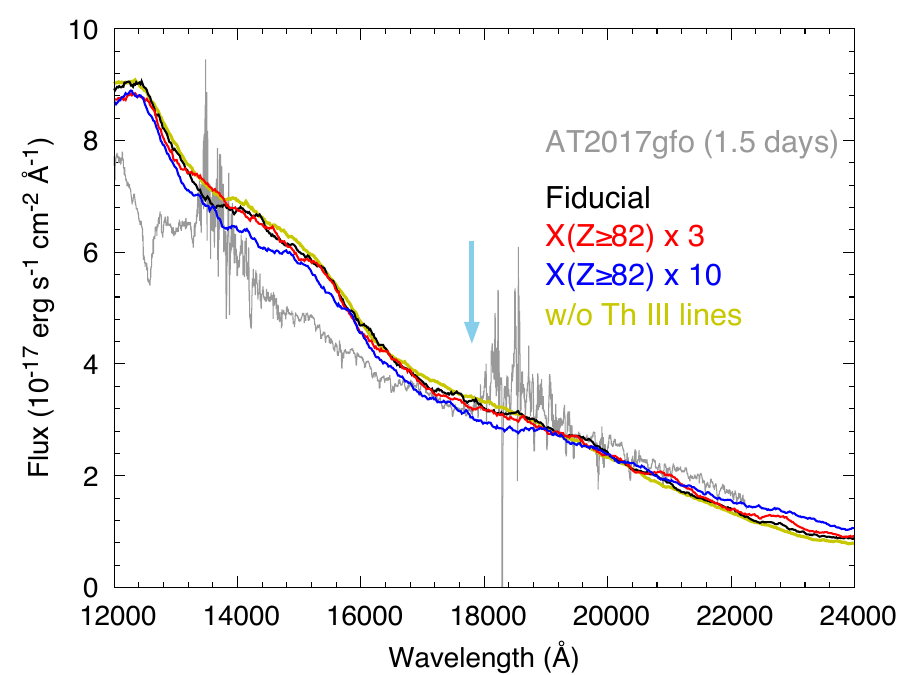}
\caption{\label{knspec} Synthetic spectra of a kilonova 1.5 days after the merger. The Black line shows the fiducial Light model results, while red and blue lines show the cases with enhanced mass fractions of the elements with $Z\ge82$ by a factor of 3 and 10, respectively, and the dark-yellow line shows the case without the Th\,{\sc iii} lines. The light-blue arrow indicates the position of Th\,{\sc iii} absorption. The observed spectrum of AT2017gfo at 1.5 days after the merger \citep[gray line,][]{pian17, smartt17} is also compared.}
\end{figure}

Fig. \ref{knspec} shows the synthetic spectra.
The black line shows the spectra calculated with the fiducial Light model.
\citet{domoto24} suggested that Th\,{\sc iii} lines can produce a broad absorption feature around 18,000 \AA.
However, our new calculations with our atomic data presented in this paper show that the appearance of the absorption features is marginal.
This is because the $gf$ values of the NIR transitions calculated in this paper tend to be lower than those adopted in \citet{domoto24} as shown in Fig. \ref{gfvs}.

The red and blue lines in Fig. \ref{knspec} show the synthetic spectra with the enhanced mass fraction of heaviest elements ($Z\ge82$), including Th.
If the mass fraction is enhanced by a factor of 3-10 as compared with our fiducial Light model (mass fraction of $(3-10) \times 10^{-5}$), kilonova spectra can show recognizable absorption features.
Our calculations demonstrate the importance of accurate atomic data in connecting the observed depth of absorption features with the actual abundance in the ejecta.

\section{Summary and conclusions}
\label{sec:conclusions}
A total of 63 energy levels were calculated and analyzed. 
In this study, the accuracy of the calculated energy levels was evaluated by comparing them with the experimental data. 
Our calculated energy levels matched the experimental data by RMS 436 cm$^{-1}$ (two levels were excluded). 
Atomic states of the $\mathrm{5f7p}$ configuration are more mixed 
than those of other configurations, and the identification 
is more complicated in $LS-$coupling. 
States of this configuration are much more pure in $JJ-$coupling.  

In this work, 455 E1-type transitions between the energy levels of the configurations considered have been calculated. To evaluate the uncertainties of the calculated line strengths,
the QQE method was 
employed, as described in references ~\cite{Pavel_Ce,Pavel_Pr,Kitoviene_Ge}. 
According to the QQE method, 46 percent fall into the D and better accuracy class. 
Our computed line strengths agree well with data computed by \citet{Safronova},  
and even better with results by \citet{Biemont}. 

Finally, we showed that kilonova spectra can show recognizable absorption features of Th\,{\sc iii} lines with a Th abundance of $(3-10) \times 10^{-5}$.
Our work demonstrates the importance of accurate atomic data in determining the mass fraction of heavy elements from observed absorption features.

\section*{Acknowledgments}
This project has received funding from the Research 
Council of Lithuania (LMTLT), agreement No S-LJB-23-1 and JSPS Bilateral Joint Research Project (JPJSBP120234201).
\section*{DATA AVAILABILITY}
The data (Table \ref{Transition}) underlying this article are available in the article and its online supplementary material.

\bibliographystyle{mnras}

\bsp	
\label{lastpage}
\end{document}